\documentclass[a4paper]{article}
\usepackage{graphicx}
\usepackage{amsmath}
\usepackage{amssymb}
\newcommand{\haak}[1]{\left(#1\right)}
\newcommand{\rhaak}[1]{\left [#1\right]}
\newcommand{\lhaak}[1]{\left | #1\right |}
\newcommand{\ahaak}[1]{\left\{#1\right\}}
\newcommand{\gem}[1]{\left\langle #1\right\rangle}

\newcommand{\rhaakl}[1]{\left[#1\right.}
\newcommand{\rhaakr}[1]{\left.#1\right]}

\newcommand{\half}{\frac{1}{2}}
\newcommand\stfm{staggered F-model}
\newcommand\tr{\operatorname{Tr}}

\begin{document}
\title{Expansions about Free-Fermion Models}

\author{Saibal Mitra\\
{\it saibalm@science.uva.nl}\\
Instituut voor Theoretische Fysica\\
Universiteit van Amsterdam\\
1018 XE Amsterdam
The Netherlands}
\date{\today}
\maketitle
\begin{abstract}
A simple technique for expanding the free energy of general six-vertex models about
free-fermion points is introduced. This technique is used to verify a Coulomb gas
prediction about the behavior of the leading singularity in the free energy of the \stfm\ at
zero staggered field.
\end{abstract}

\section{Definition of the \stfm\ }
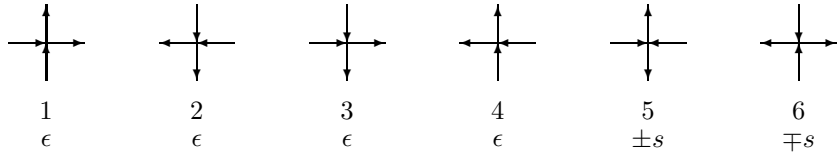
\begin{figure}
\setlength{\unitlength}{0.0165 \textwidth}
\begin{picture}(60,10)
\put(5,7.5){\vector(1,0){2.5}}
\put(2.5,7.5){\vector(1,0){2.5}}
\put(5,5){\vector(0,1){2.5}}
\put(5,7.5){\vector(0,1){2.5}}
\put(5,3){\makebox(0,0){1}}
\put(5,1){\makebox(0,0){$\epsilon $}}

\put(17.5,7.5){\vector(-1,0){2.5}}
\put(15,7.5){\vector(-1,0){2.5}}
\put(15,10){\vector(0,-1){2.5}}
\put(15,7.5){\vector(0,-1){2.5}}
\put(15,3){\makebox(0,0){2}}
\put(15,1){\makebox(0,0){$\epsilon $}}

\put(25,7.5){\vector(1,0){2.5}}
\put(22.5,7.5){\vector(1,0){2.5}}
\put(25,10){\vector(0,-1){2.5}}
\put(25,7.5){\vector(0,-1){2.5}}
\put(25,3){\makebox(0,0){3}}
\put(25,1){\makebox(0,0){$\epsilon $}}

\put(37.5,7.5){\vector(-1,0){2.5}}
\put(35,7.5){\vector(-1,0){2.5}}
\put(35,5){\vector(0,1){2.5}}
\put(35,7.5){\vector(0,1){2.5}}
\put(35,3){\makebox(0,0){4}}
\put(35,1){\makebox(0,0){$\epsilon $}}

\put(47.5,7.5){\vector(-1,0){2.5}}
\put(42.5,7.5){\vector(1,0){2.5}}
\put(45,7.5){\vector(0,1){2.5}}
\put(45,7.5){\vector(0,-1){2.5}}
\put(45,3){\makebox(0,0){5}}
\put(45,1){\makebox(0,0){$\pm s $}}

\put(55,7.5){\vector(-1,0){2.5}}
\put(55,7.5){\vector(1,0){2.5}}
\put(55,10){\vector(0,-1){2.5}}
\put(55,5){\vector(0,1){2.5}}
\put(55,3){\makebox(0,0){6}}
\put(55,1){\makebox(0,0){$\mp s $}}
\end{picture}
\setlength{\unitlength}{1 pt}
\caption{\small The six vertices and their energies. The upper and lower
signs correspond to sublattice A resp.\ sublattice B.}\label{vrt}
\end{figure}

The \stfm\ is a special case of the six-vertex model.
The six-vertex model can be defined as follows:
place arrows on the edges of a square lattice so that there are two arrows
pointing into each vertex. Six types of vertices can arise (hence 
the name of the model). These vertices are shown in fig.\ \ref{vrt}.
By giving each vertex-type a (position-dependent) energy the model is defined.
These models were first introduced to study (anti-)ferroelectric systems. Later it
was shown that six-vertex models can be mapped to solid-on-solid models
\cite{vby}.
Only a few of these models can be solved exactly. These include the
free-fermion models \cite{fan,wu} and models that can be solved using the 
Bethe Ansatz \cite{bxt2,lieb1,lieb2,lieb3,bxt1}. To define
the \stfm, we divide the lattice into two sublattices A and B, such that the
nearest neighbor of an A vertex is a B vertex. The vertex energies are chosen
as indicated in fig.\ \ref{vrt}. When the staggered field ($ s $) vanishes the model
reduces  
to the F-model, which has been solved by Lieb \cite{lieb2}. 
At zero staggered field the model is critical. In this case the groundstate is
twofold degenerate
consisting of vertices of type 5 on sublattice A and vertices of type 6 on
sublattice B,
or vice versa. For $\beta\epsilon>0$ a nonzero staggered field lifts this
degeneracy, and forces the model into
an ordered state \cite{njs}.

\section{Coulomb gas results}\label{cg}
By assuming that the F-model renormalizes to the Gaussian model, it is possible to
find the behavior
of the \stfm\ in infinitesimal staggered fields \cite{erk}. It is found that the
leading
singularity in the free energy is:
\begin{equation}\label{ssn1}
F_{s}\haak{\beta\epsilon,\beta s}\approx\haak{\beta
s}^{\frac{2}{2-\frac{\pi}{4j\haak{\beta\epsilon}}}}
\end{equation}

where
\begin{equation}\label{ssn2}
j\haak{\beta\epsilon}=\half\arccos\haak{1-\half\exp\haak{2\beta\epsilon}}
\end{equation}

At the point $\beta\epsilon=\half\ln\haak{2-\sqrt{2}}\approx -0.2674$ the exponent
becomes infinite.
Below this point a finite staggered field is necessary to force the model to an
ordered state. In this case the
transition to the ordered state happens via a Kosterlitz-Thouless
(KT) transition. The existence of a
line of KT transitions intersecting the point
$\haak{\beta\epsilon=\half\ln\haak{2-\sqrt{2}},\beta s=0}$
has been verified by combining the results of transfer matrix studies with scaling
arguments
\cite{erk}.

In this paper we will verify (\ref{ssn1}) by expanding around Baxter's exact solution on the line
$\beta\epsilon=\half\ln{2}$

\section{Baxter's solution of the \stfm}
Baxter has solved the \stfm\ at $\beta\epsilon=\half\ln(2)$
\cite{bxt}. Later it was found that this solution could be generalized to other
models if a certain condition involving the vertex weights is met. This condition
is called the free-fermion condition because for eight-vertex models satisfying this
condition the problem leads to a problem of noninteracting fermions in the 
S-matrix formulation. Let $w_{i}$ be the vertex weight for a vertex of type 
$i$ (see fig.\ \ref{vrt}), then the free-fermion condition for six-vertex models is:
\begin{equation}\label{fermion}
w_{1}w_{2}+w_{3}w_{4}-w_{5}w_{6}=0
\end{equation}
The weights $w_{i}$ may be chosen inhomogeneous. We now proceed by presenting
Baxter's solution of the \stfm.

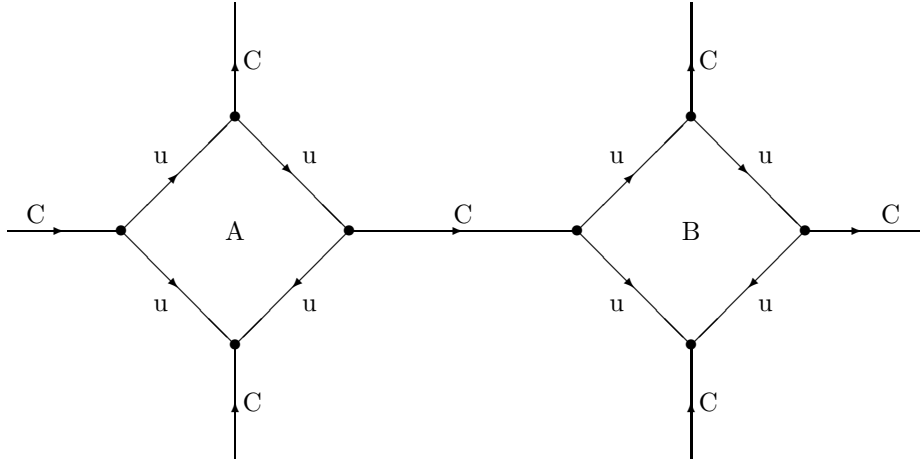
\begin{figure}
\setlength{\unitlength}{0.0625\textwidth}
\begin{picture}(16,8)
\put(1,4){\vector(1,0){0}}
\put(0,4){\line(1,0){2}}
\put(3,5){\vector(1,1){0}}
\put(2,4){\line(1,1){2}}
\put(4,7){\vector(0,1){0}}
\put(4,6){\line(0,1){2}}
\put(4,6){\line(1,-1){2}}
\put(5,5){\vector(1,-1){0}}
\put(6,4){\line(-1,-1){2}}
\put(5,3){\vector(-1,-1){0}}
\put(4,2){\line(0,-1){2}}
\put(4,1){\vector(0,1){0}}
\put(2,4){\line(1,-1){2}}
\put(3,3){\vector(1,-1){0}}
\put(0.5,4.3){\makebox(0,0){C}}
\put(2.7,5.3){\makebox(0,0){u}}
\put(5.3,5.3){\makebox(0,0){u}}
\put(5.3,2.7){\makebox(0,0){u}}
\put(2.7,2.7){\makebox(0,0){u}}
\put(4.3,1){\makebox(0,0){C}}
\put(4.3,7){\makebox(0,0){C}}
\put(2,4){\makebox(0,0){$\bullet$}}
\put(4,6){\makebox(0,0){$\bullet$}}
\put(6,4){\makebox(0,0){$\bullet$}}
\put(4,2){\makebox(0,0){$\bullet$}}
\put(11,5){\vector(1,1){0}}
\put(10,4){\line(1,1){2}}
\put(12,7){\vector(0,1){0}}
\put(12,6){\line(0,1){2}}
\put(12,6){\line(1,-1){2}}
\put(13,5){\vector(1,-1){0}}
\put(14,4){\line(-1,-1){2}}
\put(13,3){\vector(-1,-1){0}}
\put(12,2){\line(0,-1){2}}
\put(12,1){\vector(0,1){0}}
\put(10,4){\line(1,-1){2}}
\put(11,3){\vector(1,-1){0}}
\put(6,4){\line(1,0){4}}
\put(8,4){\vector(1,0){0}}
\put(14,4){\line(1,0){2}}
\put(15,4){\vector(1,0){0}}
\put(10.7,5.3){\makebox(0,0){u}}
\put(13.3,5.3){\makebox(0,0){u}}
\put(13.3,2.7){\makebox(0,0){u}}
\put(10.7,2.7){\makebox(0,0){u}}
\put(12.3,1){\makebox(0,0){C}}
\put(12.3,7){\makebox(0,0){C}}
\put(8,4.3){\makebox(0,0){C}}
\put(15.5,4.3){\makebox(0,0){C}}
\put(4,4){\makebox(0,0){A}}
\put(12,4){\makebox(0,0){B}}
\put(10,4){\makebox(0,0){$\bullet$}}
\put(12,6){\makebox(0,0){$\bullet$}}
\put(14,4){\makebox(0,0){$\bullet$}}
\put(12,2){\makebox(0,0){$\bullet$}}
\end{picture}
\setlength{\unitlength}{1 pt}
\caption{\small The ''cities'' on the decorated lattice. A and
B refer to the two sublattices. The meaning of the orientations on the edges is
explained in the text.}\label{dm}
\end{figure}

Divide the lattice into two sublattices A and B. Choose the vertex energies as
indicated
in fig.\ \ref{vrt}. Consider the ground state in which all A-vertices are vertices
of type 6,
and all B-vertices are of type 5. Any state can now be represented by drawing
lines on the lattice where the arrows point oppositely to the ground state
configuration.
In terms of these lines the six vertices are represented by vertices with either
no lines, two lines at right angles, or four lines. The energies of these vertices
are respectively $-s$, $\epsilon$ and $s$. The next step is to replace the original
lattice
by a decorated lattice by replacing each original vertex by a ''city'' of four
internally
connected points (see fig.\ \ref{dm}).

The lines on the original lattice are regarded as dimers on the external edges
of the decorated lattice. For any configuration on the original lattice, it is
possible to place dimers on the internal edges of the decorated lattice, so that
the lattice becomes completely covered. Now associate to each dimer a weight as
indicated in fig.\ \ref{dm}. Demanding that the closed-packed
dimer problem formulated on the decorated lattice is equivalent to our original 
problem yields:
\begin{eqnarray}
C&=&\exp\haak{-\half\beta s}\\
u&=&\half\sqrt{2}\exp\haak{\half\beta s}\\
\beta\epsilon&=&\half\ln\haak{2}\label{ferm}
\end{eqnarray}

To solve the close-packed dimer problem, we use the Pfaffian method \cite{hrst, kst,
mnt}.
This method expresses the partition function $Z$ for a closed packed dimer model
on an $N$ by $M$ planar lattice:
\begin{equation}\label{pfaff}
Z^{2}=\det R
\end{equation}
Here $R$ is an $N\times M$ by $N\times M$ anti-symmetric matrix, defined as follows.
Enumerate all the $N\times M$ vertices on the decorated lattice.
If vertex i is not connected to vertex j via an edge,  $R_{i,j}=0$, else $R_{i,j}=\pm$ 
fugacity of dimer at edge connecting i to j.
The way the signs have to be chosen is explained \cite{kst}. These signs define an
orientation of the
edges. Positive $R_{i,j}$ is indicated by an arrow pointing from $i$ to $j$.

To set up a perturbation theory about $\beta\epsilon=\half\ln\haak{2}$, we also
need the inverse of $R$. Both the determinant and the inverse of $R$ are easily
calculated by performing a similarity transformation, see \cite{bxt} for details.
The determinant yields the following expression for the reduced free energy per
vertex (i.e.\ the free energy times $-\beta$), 
denoted as $F_{\text{Baxter}}$, for an infinite by infinite lattice:

\begin{equation}\label{frbx}
\begin{array}{l}
F_{\text{Baxter}}=\lim_{N,M\rightarrow\infty}\frac{1}{2NM}\ln\det R\\
=\frac{1}{8\pi^{2}}\int_{0}^{2\pi}\!\int_{0}^{2\pi}
\ln\rhaak{2\cosh\haak{2\beta s}+2\cos\haak{\theta_{1}}\cos\haak{\theta_{2}}}
d\theta_{1}d\theta_{2}\\
\end{array}
\end{equation}

\section{Perturbation theory}\label{pbth}
We now proceed with the derivation of a perturbation theory about the free-fermion
line of a six-vertex model. The Hamiltonian of a general six-vertex model can be
defined as
follows. One assigns an energy $e\haak{p,i}$ to a vertex in state $p$ (see fig.\
\ref{vrt}) and position $i$. The configuration of the lattice can be specified by a
function $c$ which maps a position of a vertex to a number, $1\cdots 6$, which 
is to be interpreted as the state of the vertex at that position. The reduced
Hamiltonian ($H$) is defined to be the functional that assigns to each state
$c$ its energy times $-\beta$. We can thus write
\begin{equation}
H\haak{c}=-\beta\sum_{i}e\haak{c\haak{i},i}
\end{equation}

For $H$ a Hamiltonian of a general six-vertex model and $H_{0}$ a 
Hamiltonian of a free-fermion model, a perturbation $V$ can be defined so that
we have
\begin{equation}
H=H_{0}+V
\end{equation}
The partition function $Z$ can be written as:
\begin{equation}
Z=\sum_{c}\exp\haak{H_{0}\haak{c}+V\haak{c}}=Z_{0}\gem{\exp\haak{V}}
\end{equation}
Here $Z_{0}$ is the partition function
of the free-fermion model. The reduced free energy can be expressed as:
\begin{equation}\label{fermcum}
F=F_{0}+\ln\gem{\exp\haak{V}} =F_{0}+\gem{V}+\half\gem{\haak{V-\gem{V}}^{2}}+\ldots
\end{equation}
Here $F_{0}$ is the reduced free energy of the free-fermion model. Now write
$V=\sum_{i}V_{i}$
with $V_{i}\haak{c\haak{i}}$ a perturbation of the vertex energy times $-\beta$ at
position $i$. (\ref{fermcum}) can
be rewritten as:
\begin{equation}\label{frmcm}
F=F_{0}+\sum_{i}\gem{V_{i}}+\half\sum_{ij}\rhaak{\gem{V_{i}V_{j}}-\gem{V_{i}}\gem{V_{j}}}+\ldots
\end{equation}
To compute a free-fermion average $\gem{V_{i_{1}}V_{i_{2}}\ldots V_{i_{n}}}$, we
can proceed as follows: Introduce a constraint in the free-fermion model by
requiring the vertices at the positions $i_{1}\ldots i_{n}$ to be in the states 
$x_{1}\ldots x_{n}$. The
partition function of this model is denoted by $Z_{i_{1}\cdots
i_{n}}\haak{x_{1}\ldots x_{n}}$. 
We can then write
\begin{equation}\label{gm}
\gem{V_{i_{1}}V_{i_{2}}\ldots V_{i_{n}}}=\sum_{x_{1}\ldots
x_{n}}\frac{Z_{i_{1}\cdots i_{n}}\haak{x_{1}\ldots x_{n}}
V\haak{x_{1}}\ldots V\haak{x_{n}}}{Z_{0}}
\end{equation}
It now remains to calculate $Z_{i_{1}\cdots i_{n}}\haak{x_{1}\ldots x_{n}}$.
It is convenient to reformulate this problem as follows: Denote the state of
an arrow located at the edge $j$ by $s_{j}$. Put $s_{j}=1$ if the arrow points
oppositely to the ground state configuration and $s_{j}=0$ otherwise. Define
a constrained free-fermion model by requiring the arrow at the edge $j_r$ to
be in state $s_{j_{r}}$ for $1\leq r\leq m$. We then want to evaluate the partition
function of 
this model, which we denote as $Z^{\text{cons}}\haak{s_{j_{1}}\ldots s_ {j_{m}}}$. The idea is to
perturb the weights of the dimers on the edges $j_{r}$ infinitesimally. We redefine
the weight of the dimer on the edge $j_{r}$ by multiplying it by
$\haak{1+\epsilon_{r}}$. The partition 
function of the redefined free-fermion model
($Z\haak{\epsilon_{1}\ldots\epsilon_{m}}$) 
can be written in terms of the constrained partition functions as:
\begin{equation}\label{part1}
\begin{split}
& Z\haak{\epsilon_{1}\ldots\epsilon_{m}}=\sum_{\ahaak{s}}Z^{\text{cons}}\haak{s_{j_{1}}\ldots
s_{j_{m}}}\prod_{k=1}^{m}\haak{1+s_{j_{k}}\epsilon_{k}}\\
& =Z_{0}+\sum_{k}Z^{\text{cons}}\haak{s_{j_{k}}=1}\epsilon_{k}+\sum_{k<l}Z^{\text{cons}}\haak{s_{j_{k}}=1
,s_{j_{l}}=1}\epsilon_{k}\epsilon_{l}+\ldots
\end{split}
\end{equation}
$Z\haak{\epsilon_{1}\ldots\epsilon_{m}}$ can be calculated using (\ref{pfaff}),
by making the necessary changes to $R$. We can write:
\begin{equation}\label{mtrx}
R=R_{0}+\sum_{k=1}^{m}\epsilon_{k}R_{\haak{k}}
\end{equation}
Here $R_{0}$ is the original unperturbed matrix, $R_{\haak{k}}$ is defined as follows:
\begin{eqnarray*}
&R_{\haak{k},ij}=R_{0,ij}\\ 
&\mbox{if $i$ and $j$ are connected by $j_{k}$}\\
&R_{\haak{k},ij}=0\\ 
&\mbox{if $i$ and $j$ are not connected by $j_{k}$.}
\end{eqnarray*}
Note that the $R_{\haak{k}}$ have only two nonzero matrix elements. Inserting
(\ref{mtrx}) in (\ref{pfaff}) and expanding gives:
\begin{equation}\label{zedeps}
\begin{split}
& Z\haak{\epsilon_{1}\ldots\epsilon_{m}}=\sqrt{\det R}=\sqrt{\det
R_{0}}\exp\haak{\half\tr\ln\rhaak{1+\sum_{k}\epsilon_{k}R_{0}^{-1}R_{\haak{k}}}}\\
& =\sqrt{\det R_{0}}\rhaakl{1+\half\sum_{k}\epsilon_{k}\tr\haak{R_{0}^{-1}R_{\haak{k}}}}\\
& \rhaakr{\mbox{}+\frac{1}{4}\sum_{k,l}\epsilon_{k}\epsilon_{l}\rhaak{\half\tr\haak{R_{0}^{-1}R_{\haak{k}}}
\tr\haak{R_{0}^{-1}R_{\haak{l}}}-\tr\haak{R_{0}^{-1}R_{\haak{k}}
R_{0}^{-1}R_{\haak{l}}}}+\ldots}
\end{split}
\end{equation}
Using (\ref{zedeps}) and (\ref{part1}) we can directly read off the constrained
partition functions if all the
constrained arrows point oppositely to the ground state configuration. To calculate
a general constrained partition function one can apply
the principle of inclusion and exclusion. E.g.\ consider the
evaluation
of $Z\haak{s_{1},s_{2},s_{3},s_{4},s_{5}}$, with $s_{1}=s_{2}=1$ and 
$s_{3}=s_{4}=s_{5}=0$. Put $t_{3}=t_{4}=t_{5}=1$. According to the principle 
of incusion and exclusion, we can write:
\begin{eqnarray}
Z\haak{s_{1},s_{2},s_{3},s_{4},s_{5}}&=&Z\haak{s_{1},s_{2}}\nonumber\\
&&\mbox{}-\rhaak{Z\haak{s_{1},s_{2},t_{3}}+
Z\haak{s_{1},s_{2},t_{4}}+Z\haak{s_{1},s_{2},t_{5}}}\nonumber\\
&&\mbox{}+Z\haak{s_{1},s_{2},t_{3},t_{4}}+Z\haak{s_{1},s_{2},t_{3},t_{5}}+Z\haak{s_{1},s_{2},t_{4},t_{5}}\nonumber\\
&&\mbox{}-Z\haak{s_{1},s_{2},t_{3},t_{4},t_{5}}\label{minv}
\end{eqnarray}
\section{First order computation for the \stfm\ }
For the \stfm\ the expansion can be simplified. The vertex in the ground state 
at a particular point will be referred to as an a-vertex. A b-vertex is obtained
by reversing the arrows of an a-vertex. An a-vertex (b-vertex) is thus of type 
5 or 6 and has an energy of $-s$ $\haak{s}$. The constrained partition function 
corresponding to the model with one vertex constrained to be an a-vertex (b-vertex)
is denoted as $Z_{\text{a}}$ $\haak{Z_{\text{b}}}$. Note that under the transformation
$s\rightarrow -s$ 
the role of vertices a and b are interchanged. We thus have
\begin{equation}\label{sym}
Z_{\text{a}}\haak{\beta s}=Z_{\text{b}}\haak{-\beta s}
\end{equation}
If we put $\beta\epsilon=\half\ln\haak{2}+U$ we have, according to (\ref{frmcm})
and (\ref{gm}), to first order in $U$:
\begin{equation}\label{fu}
F=F_{0}-\frac{Z_{0}-Z_{\text{a}}-Z_{\text{b}}}{Z_{0}}U+O\haak{U^{2}}
\end{equation}
Here $F$ is the reduced free energy per vertex of the \stfm, and $F_{0}=F_{\text{Baxter}}$ in
(\ref{frbx}). To calculate $Z_{\text{b}}$ we
only have to constrain two opposing arrows of one vertex to point oppositely
to an a-vertex. Using the formalism of the previous section, we have obtained:
\begin{equation}\label{wb}
\frac{Z_{\text{b}}}{Z_{0}}=\frac{1}{64\pi^{4}}\rhaak{\int_{0}^{2\pi}\int_{0}^{2\pi}d\theta_{1}d\theta_{2}
\frac{\exp\haak{-2\beta
s}+\cos\haak{\theta_{1}}\cos\haak{\theta_{2}}}{\cosh\haak{2\beta s}
+\cos\haak{\theta_{1}}\cos\haak{\theta_{2}}}}^{2}
\end{equation}
Using this, the first order expansion of the free energy can be written as:

\begin{equation}\label{ou}
F=F_{0}+\half\rhaak{\haak{\frac{\partial F_{0}}{\partial\beta s}}^{2}-1}U+\ldots
\end{equation}
\section{Singular behavior in the vicinity of the free-fermion line}
We will now verify the Coulomb gas result (see section \ref{cg}):
\begin{equation}\label{sn1}
F_{s}\sim\haak{\beta s}^{\frac{2}{2-\frac{\pi}{4j\haak{\beta\epsilon}}}}
\end{equation}
where 
\begin{equation}\label{sn2}
j\haak{\beta\epsilon}=\half\arccos\haak{1-\half\exp\haak{2\beta\epsilon}}
\end{equation}
If we put 
\begin{equation}\label{jeps}
\beta\epsilon=\half\ln\haak{2}+U
\end{equation}
Expanding in powers of $U$ yields:
\begin{equation}
F_{s}=A\haak{U}\haak{\beta
s}^{2}\rhaak{-\frac{8}{\pi}\haak{U+O\haak{U^{2}}}\ln\haak{\beta s}
+\frac{32}{\pi^{2}}\haak{U^{2}+O\haak{U^{3}}}\ln^{2}\lhaak{\beta s}+\ldots}
\end{equation}
where the amplitude $A\haak{U}$ is a meromorphic function.
If we compare this with the non-analytical behavior at $U=0$ (see (\ref{sng}) in the
appendix),
we find:
\begin{equation}
A\haak{U}=\frac{1}{4U} +O\haak{1}
\end{equation}
It then follows that the amplitude of the term $\haak{\beta
s}^{2}\ln^{2}\lhaak{\beta s}$ is
$\frac{8}{\pi^{2}}\haak{U+O\haak{U^{2}}}$. It is now a simple matter to verify
this using (\ref{ou}) and (\ref{sng}). From (\ref{sng}) and (\ref{ou}) it follows
that the order $U$ contribution to the singular part of the reduced free energy,
$F_{1}\haak{\beta s}$, can be written as
\begin{equation}
F_{1,s}\haak{\beta s}=\rhaak{B_{1}\haak{\beta s}\ln\lhaak{\beta s}+B_{2}\haak{\beta
s}\ln^{2}\lhaak{\beta s}}U
\end{equation}
with $B_{1}$ and $B_{2}$ regular functions of $\beta s$. Inserting (\ref{sng})
in (\ref{ou}) gives 
\begin{equation}
B_{2}\haak{\beta s}=\frac{8}{\pi^{2}}\rhaak{\haak{\beta
s}^{2}-\frac{2}{3}\haak{\beta s}^{4}
+\frac{79}{90}\haak{\beta s}^{6}+\ldots}
\end{equation}
We have thus verified (\ref{sn2}) to first order in $U$.

\section{Conclusions and outlook}
We have presented a simple technique for expanding the free energy of six-vertex
models about
free-fermion points. Applying this technique to the \stfm\ has enabled us to verify a
Coulomb gas prediction about the singular part of the free
energy of this model. It would be interesting to perform such computations to higher order in the
free-fermion expansion.
It is possible that such an undertaking might lead to proofs of certain Coulomb gas
results.

\appendix
\section{Singular part of the free energy}
In this appendix we calculate the singular part of the free energy of the
\stfm\ at $\beta s=0$ on the free-fermion line. Expanding the logarithm in (\ref{frbx}) yields
\begin{equation}
\begin{array}{ccc}
F_{\text{Baxter}}\haak{\beta
s}&=&-\frac{1}{8\pi^{2}}\int_{-\pi}^{\pi}\int_{-\pi}^{\pi}d\theta_{1}d\theta_{2}
\sum_{n=1}^{\infty}\frac{\cos^{n}\haak{\theta_{1}}\cos^{n}\haak{\theta_{2}}}
{n\cosh^{n}\haak{2\beta s}}\\
&=&-\half\sum_{n=1}^{\infty}\frac{1}{2n\cosh^{2n}\haak{2\beta s}}
\rhaak{\frac{\haak{2n}!}{4^{n}n!^{2}}}^{2}
\end{array}
\end{equation}
Using the asymptotic expansion
\begin{equation}
n!=n^{n}\exp\haak{-n}\sqrt{2\pi
n}\exp\haak{{\sum_{k=1}^{\infty}\frac{B_{2k}}{2k\haak{2k-1}}
\frac{1}{n^{2k-1}}}}
\end{equation}
where the $B_{r}$ are the Bernoulli numbers, we find
\begin{equation}\label{fex}
F_{\text{Baxter}}\haak{\beta
s}=-\frac{1}{4\pi}\sum_{n=1}^{\infty}\frac{1}{n^{2}\cosh^{2n}\haak{2\beta s}}
\rhaak{1-\frac{1}{4n}+\frac{1}{32n^{2}}+\frac{1}{128n^{3}}+\cdots}
\end{equation}
We can find the non-analytical part of the function $\sum_{n=1}^{\infty}\frac{1}
{n^{p}\cosh^{2n}\haak{2\beta s}}$ as follows:
Put $t=\ln\haak{\cosh^{2}\haak{2\beta s}}$. We then have to find the non-analytical part
of the function $U_{p}\haak{t}$ with
\begin{equation}\label{udf}
U_{p}\haak{t}=\sum_{n=1}^{\infty}\frac{\exp\haak{-n t}}{n^{p}}
\end{equation}
at $t = 0$ for $p\geq 2$. From (\ref{udf}) it follows that 
\begin{equation}\label{rec}
\frac{dU_{p+1}}{dt}=-U_{p}
\end{equation}
We denote the non-analytical part of $U_{p}$ by $\tilde{U}_{p}$. It then follows from
(\ref{rec}) that
\begin{equation}\label{recs}
\frac{d\tilde{U}_{p+1}}{dt}=-\tilde{U}_{p}
\end{equation}
For $p=1$ the sum in (\ref{udf}) is easily evaluated:
\begin{equation}
U_{1}\haak{t}=-\ln\haak{1-\exp\haak{-t}}
\end{equation}
And we see that $\tilde{U}_{1}\haak{t}$ is given by
\begin{equation}\label{u1}
\tilde{U}_{1}\haak{t}=-\ln\haak{t}
\end{equation}
From (\ref{u1}) and (\ref{recs}) it then follows that
\begin{equation}
\tilde{U}_{p}\haak{t}=\haak{-1}^{p}\frac{t^{p-1}}{\haak{p-1}!}\ln\haak{t}
\end{equation}
Inserting this in (\ref{fex}) gives 
\begin{equation}\label{fst}
F_{s}\haak{\beta s}=-\frac{1}{4\pi}\haak{t+\frac{t^{2}}{8}+\frac{t^{3}}{192}
-\frac{t^{4}}{3072}+\cdots}\ln\haak{t}
\end{equation}
Where $F_{s}\haak{\beta s}$ is the singular part of the free energy and 
$t=2\ln\haak{\cosh\haak{2\beta s}}$. Expanding (\ref{fst}) in powers of $\beta s$ gives
\begin{equation}\label{sng}
F_{s}\haak{\beta s}=-\frac{2}{\pi}\rhaak{\haak{\beta s}^{2}-\frac{1}{6}\haak{\beta
s}^{4} 
+\frac{23}{180}\haak{\beta s}^{6}-\frac{593}{5040}\haak{\beta s}^{8}+\cdots}
\ln\lhaak{\beta s}
\end{equation}


\begin{thebibliography}{99}
\bibitem{bxt}Baxter R J 1970 Phys.\ Rev.\ B {\bf 1} 2199.
\bibitem{bxt1}Baxter R J 1971 Stud.\ Appl.\ Math.\ (M.I.T.) {\bf 50} 51
\bibitem{bxt2}Baxter R J 1972 Ann.\ Phys.\ {\bf 70} 193.
\bibitem{vby}van Beijeren H 1977 Phys.\ Rev.\ lett.\ {\bf 38} 993.
\bibitem{fan}Fan C and Wu F Y 1970 Phys.\ Rev.\ B {\bf 2} 723.
\bibitem{hrst}Hurst C A and Green H S 1960 J.\ Chem.\ Phys.\ {\bf 33} 1059.
\bibitem{kst}Kasteleyn P W 1963 J.\ Math.\ Phys.\ {\bf 4} 287.
\bibitem{lieb1}Lieb E H 1967 Phys.\ Rev.\ {\bf 162} 162.
\bibitem{lieb2}Lieb E H 1967 Phys.\ Rev.\ Lett.\ {\bf 18} 1046.
\bibitem{lieb3}Lieb E H 1967 Phys.\ Rev.\ Lett.\ {\bf 18} 108.
\bibitem{erk}Luijten E, van Beijeren H, Bl\"{o}te H W J 1994 Phys.\ Rev.\ Lett.\
{\bf 73}
456.
\bibitem{mnt}Montroll E W 1964 Combinatorial Mathematics, John Wiley \& Sons, inc.,
New York.
\bibitem{njs} den Nijs M 1979 J.\ Phys.\ A {\bf 12} 1857.
\bibitem{wu}Wu F Y and Lin K Y 1975 Phys.\ Rev.\ B {\bf 12} 419.
\bibitem{jeps}Youngblood R W and Axe J D 1980 Phys.\ Rev.\ B {\bf 21} 5212.
\end{thebibliography}
\end{document}